\documentclass{elsart}

\usepackage[square,comma]{natbib}
\usepackage{graphicx}
\usepackage{pxfonts}
\usepackage{lineno}

\usepackage{amssymb}
\usepackage[hyphens]{url}

\journal{}

\begin{document}

\thispagestyle{empty}
\begin{Large}
\textbf{DEUTSCHES ELEKTRONEN-SYNCHROTRON}

\textbf{\large{Ein Forschungszentrum der Helmholtz-Gemeinschaft}\\}
\end{Large}

DESY 11-200

November 2011

\begin{eqnarray}
\nonumber &&\cr \nonumber && \cr \nonumber &&\cr
\end{eqnarray}
\begin{eqnarray}
\nonumber
\end{eqnarray}
\begin{center}
\begin{Large}
\textbf{Analytical studies of constraints on the performance for
EEHG FEL seed lasers}
\end{Large}
\begin{eqnarray}
\nonumber &&\cr \nonumber && \cr
\end{eqnarray}

\begin{large}
Gianluca Geloni,
\end{large}
\textsl{\\European XFEL GmbH, Hamburg}
\begin{large}

Vitali Kocharyan and Evgeni Saldin
\end{large}
\textsl{\\Deutsches Elektronen-Synchrotron DESY, Hamburg}
\begin{eqnarray}
\nonumber
\end{eqnarray}
\begin{eqnarray}
\nonumber
\end{eqnarray}
ISSN 0418-9833
\begin{eqnarray}
\nonumber
\end{eqnarray}
\begin{large}
\textbf{NOTKESTRASSE 85 - 22607 HAMBURG}
\end{large}
\end{center}
\clearpage
\newpage

\begin{frontmatter}



\title{Analytical studies of constraints on the performance for EEHG FEL seed lasers}


\author[XFEL]{Gianluca Geloni\thanksref{corr},}
\thanks[corr]{Corresponding Author. E-mail address: gianluca.geloni@xfel.eu}
\author[DESY]{Vitali Kocharyan}
\author[DESY]{and Evgeni Saldin}

\address[XFEL]{European XFEL GmbH, Hamburg, Germany}
\address[DESY]{Deutsches Elektronen-Synchrotron (DESY), Hamburg,
Germany}

\begin{abstract}
Laser seeding technique have been envisioned to produce nearly
transform-limited pulses at soft X-ray FELs. Echo-Enabled Harmonic
Generation (EEHG) is a promising, recent technique for harmonic
generation with an excellent up-conversion to very high harmonics,
from the standpoint of electron beam physics. This paper explores
the constraints on seed laser performance for reaching wavelengths
of $1$ nm. We show that the main challenge in implementing the EEHG
scheme at extreme harmonic factors is the requirement for accurate
control of temporal and spatial quality of the seed laser pulse. For
example, if the phase of the laser pulse is chirped before
conversion to an UV seed pulse, the chirp in the electron beam
microbunch turns out to be roughly multiplied by the harmonic
factor. In the case of a Ti:Sa seed laser, such factor is about
$800$. For such large harmonic numbers, generation of nearly
transform-limited soft X-ray pulses results in challenging
constraints on the Ti:Sa laser. In fact, the relative discrepancy of
the time-bandwidth product of the seed-laser pulse from the ideal
transform-limited performance should be no more than one in a
million. The generated electron beam microbunching is also very
sensitive to distortions of the seed laser wavefront, which are also
multiplied by the harmonic factor. In order to have minimal
reduction of the FEL input coupling factor, it is desirable that the
size-angular bandwidth product of the UV seed laser beam be very
close to the ideal i.e. diffraction-limited performance in the waist
plane at the middle of the modulator undulator.
\end{abstract}

%
%
%
\end{frontmatter}



\section{\label{sec:uno} Introduction}

An important goal for any advanced X-ray FEL is the production of
X-ray pulses with the minimum allowed photon energy width for given
pulse length, which defines the transform-limit. A well-known
approach to obtain fully coherent radiation in the soft X-ray region
relies on frequency multiplication, a scheme known as
high-gain-harmonic-generation (HGHG) \cite{HGHG1,HGHG2}.  In a HGHG
FEL, the radiation output is obtained from a coherent subharmonic
seed laser pulse. Consequently, the optical properties of HGHG FELs
are expected to reflect the characteristics of the high-quality seed
laser. Echo-enabled harmonic generation (EEHG) is a recent,
promising technique for efficient harmonic generation
\cite{EEHG1,EEHG2}. The key advantage of EEHG over HGHG is that the
amplitude of the achieved microbunching factor decays slowly with an
increasing harmonic number. Consequently, as concerns electron beam
physics issues, EEHG allows for the generation of fully coherent
radiation at soft X-ray wavelengths with a single upshift stage, and
using a conventional optical laser system. The remarkable
up-frequency conversion efficiency of the method has stimulated wide
interest to generate near transform-limited soft X-ray pulses.
Several EEHG FEL projects are now under development
\cite{LCLS2E}-\cite{FLEEHG}. A typical EEHG setup consists of two
stages for electron beam phase space manipulation, followed by a
radiator. Each stage includes an undulator, which is used to
modulate the electron beam in energy with the help of a seed laser,
and a chicane following the modulator, which is used to apply
energy-dependent slippage to the electrons. The radiator is composed
by a sequence of undulators tuned to the desired output wavelength.
This final section is similar to that used for SASE FELs. However it
is shorter, and produces coherent radiation only because the beam
has been coherently prebunched. The seed laser is assumed to be
tuned at $200$ nm (or $270$ nm) corresponding to the fourth harmonic
(or the third) of a Ti:Sapphire laser.

Limitations on the performance of EEHG schemes related with electron
beam dynamics issues, as the beam goes through the various
undulators and chicanes, has been extensively discussed in
literature \cite{EETOL1,EETOL2,EETOL3} and goes beyond the scope of
this paper. Here we will focus our attention on the first part of
the harmonic generation process, discussing the constraints on the
seed laser performance needed for reaching wavelengths of about $1$
nm. In fact, in chirped-pulse amplification systems (CPA) systems,
both temporal and spatial quality of the beam can be degraded due to
the propagation through the optical components, non-linear effects
or inhomogeneous doping concentration in the amplifying media, and
thermal effects linked to the pumping process. In particular, the
aim of this work is to evaluate the impact of variations of the
characteristics of output radiation when the FEL is seeded by a
laser with non-ideal properties, including effects such as linear
and nonlinear frequency chirp and wavefront distortion.

A description of the impact of phase chirp of the EEHG (or HGHG) FEL
output can be made without numerical simulation codes. In fact, as
is well-known, if the phase of the seed laser is chirped, the chirp
is simply multiplied by the frequency multiplication factor $N$. In
this case, the method used to describe the output field perturbation
is independent of the specific kind of harmonic generation
technique: it only depends on the frequency multiplication factor
$N$. It follows that both EEHG and HGHG FELs starting from a Ti:Sa
laser with a wavelength around $800$ nm can produce
transform-limited radiation down to wavelengths of $1$ nm only when
the relative discrepancy of the time-bandwidth product of the
compressed $800$ nm laser pulse from the ideal, transform-limited
performance is no more than one in a million, roughly corresponding
to the squared of the harmonic number. However, pulses from a
commercially available Ti:Sapphire chirped pulse amplifiers are
usually limited in such discrepancy to around $1\%$, due to
non-ideal effects. Therefore, research and development activities
must be performed in order to reach the required temporal quality.
To the best of our knowledge, there is only one article\footnote{The
issue was also discussed during the preparation of this work in
\cite{RATN}. There HGHG was mainly considered but, as noted above,
the method used to describe the output field perturbation is
independent of the specific kind of harmonic generation technique.}
reporting on the impact of temporal variations in seed laser pulses
on EEHG FEL output radiation characteristics, \cite{EETOL3}. The
analysis in \cite{EETOL3} is based on numerical simulations in the
case of $1.2$ nm output radiation wavelength. According to results
of the sensitivity study in \cite{EETOL3}, the phase of the $202$ nm
seed laser pulse (corresponding to the fourth harmonic of a Ti:Sa
laser) must to be controlled to within $0.5$ degrees. Consequently,
the phase of the Ti:Sa laser output must to be controlled to within
roughly $0.1$ degrees. This result is consistent with our analysis,
which is performed at a very elementary level.

We also evaluate the impact on the EEHG FEL output of wavefront
errors in the seed laser. In the case of ideal performance, the seed
(UV) laser beam must be characterized by a flat (i.e.
diffraction-limited) wavefront in the waist plane in the middle of
the modulator undulator. If the wavefront exhibits errors, errors in
the microbuch wavefront follow, which are multiplied by the
frequency multiplication factor. These microbunch wavefront errors
do not affect the spatial quality of the FEL output radiation, which
is the same for both perturbed and unperturbed wavefront cases. They
only affect the input signal value at the target harmonic. However,
because of the exponential dependence of the signal suppression
factor on the wavefront errors, one obtains an appreciable FEL
output only when phase errors are sufficiently small to give
appreciable input signal. As a result, the seed UV laser beam must
exhibit a nearly diffraction-limited wavefront in the waist plane,
with very little phase variation. In particular for a target
harmonic with wavelengths of about $1$ nm, the wavefront of the UV
beam must be controlled to within a fraction of a degree across the
electron beam area. These relatively small phase variations cause
the signal at the entrance of the FEL amplifier to drop of a
quantity of order of the ideal (diffraction-limited) performance. In
contrast with phase variation in time, the spatial quality of UV
seed laser beam can be improved by means of active optics and
spatial filtering. However, these manipulations with laser beam
usually cause significant losses in beam power.

To the best of our knowledge, the crucially important problem of
seeding with beam wavefront distortions was only recently reported
in workshops \cite{WORK1,WORK2}, where the impact of wavefront
errors on the EEHG performance was discussed, based on numerical
simulations, in the case of the highest target harmonic at $13$ nm.
Results of \cite{WORK1,WORK2} are consistent with our analysis,
which has been performed purely analytically.

The suppression of the output signal due to phase variations in
space seems somehow in contrast with the effects of phase variations
in time, where phase errors affect the temporal quality of the
output radiation, but not the FEL output power. From this viewpoint,
it should be noted that the radiation field is characterized by
notions such as temporal and spatial coherence. The transverse
coherence of FEL radiation develops automatically, without laser
seeding. This happens due to transverse eigenmode selection: due to
different gains of the FEL transverse eigenmodes, only one survives
at the end of the FEL process. The coherence time is defined by the
inverse FEL amplification bandwidth. For conventional soft X-ray
FELs the typical amplification bandwidth is much wider than the
Fourier transform limited value corresponding to the radiation pulse
duration, meaning that the coherence time is much shorter than the
pulse duration. Consequently, microbunch phase variations in time
only lead to phase variations in the output radiation pulse, without
suppression of the output power level.

\section{\label{sec:due} Issues affecting the performance of EEHG FEL}

Phase control is an important aspect in the development of all FEL
sources based on harmonic generation. Methods for dealing with
issues concerning temporal phase variations in frequency multipliers
are based on the same general principle \cite{ROBI}:  the effect of
frequency multiplication by a harmonic factor $N$, is to multiply
the phase variation by $N$. The EEHG scheme is obviously based on
harmonic generation, but is more complicated than other schemes, and
consists of two modulators, two dispersion sections, and one
radiator undulator. A unique feature of EEHG scheme is the
utilization of two different seed laser pulses which can have
different temporal and spatial quality. It is thus natural to
investigate the question whether the general principle above can
also be applied to EEHG. Analytical results \cite{EEHG2} refer to
the specific model an of infinitely long, uniform electron bunch
only. This steady state model proved to be very fruitful, allowing
for simple analytical expressions describing the main
characteristics of EEHG scheme. However, as discussed above, the
seed laser pulses and, consequently the electron beam microbunching,
are always characterized by phase variations in time and space
(wavefront distortions). We will therefore extend analytical
description of EEHG scheme in \cite{EEHG2}, following the line of
derivations in that reference, to the time dependent case and
account for finite duration and transverse size of the electron
bunch.

To this aim, we assume that the temporal profile of the electron
beam can be modeled as a Gaussian, and that the initial electron
beam distribution can be factorized as a product of energy,
$f_{0p}(p)$, and density, $f_{0\zeta}(\zeta)$ distributions as

\begin{eqnarray}
f_0(\zeta,p)= f_{0p}(p) f_{0\zeta}(\zeta) = \frac{N_0}{2\pi
\sigma_\zeta} \exp\left[-\frac{p^2}{2}-\frac{\zeta^2}{2
\sigma_\zeta^2}\right]~. \label{f0new}
\end{eqnarray}
Here $p=(E-E_0)/\sigma_E$ is the dimensionless energy deviation of a
particle from the average energy $E_0$, and the rms spread is given
by $\sigma_E$. Similarly, $\zeta = \omega_l t$ is the dimensionless
time, with $\omega_l$ the laser frequency, assumed to be the same in
both stages, and $\sigma_\zeta$ is the rms spread of the density
distribution. Finally $N_0$ is the total number of particles in the
beam. The longitudinal phase space is described by the variables
$(\zeta,p)$. Passing through the first modulator and dispersive
section the phase space variables transform to $(\zeta',p')$, which
are given by

\begin{eqnarray}
p'=p + A_1 \sin (\zeta + \phi_1) ~~, ~~~ \zeta'=\zeta + B_1 p'~,
\label{M1}
\end{eqnarray}
where $A_1 = \Delta E_1/\sigma_E$, $\Delta E_1$ being the energy
modulation imposed by the seed laser, $\phi_1=\phi_1(\zeta)$ is the
phase of the laser pulse, which depends on the time $\zeta$, and
$B_1 = R_{56}^{(1)} \sigma_E \omega_l/(E_0 c)$, $R_{56}^{(1)}$ is
the strength of the first chicane. Substituting Eq. (\ref{M1}) into
Eq. (\ref{f0new}) one can obtain the distribution after the first
modulator and dispersive section.
%
The new phase space variables $(\zeta',p')$ will transform after the
passage through the second modulator and dispersive section, to
$(\zeta'',p'')$, which are given, in a similar way, by

\begin{eqnarray}
p''=p' + A_2 \sin (\zeta' + \phi_2)~~, ~~~ \zeta''=\zeta' + B_2
p''~, \label{M2}
\end{eqnarray}
the subscript $'2'$ referring to the second stage\footnote{We kept
our notation similar to that of \cite{EEHG2}. However, we chose
$\omega_l \equiv \omega_1=\omega_2$ from the very beginning.
Therefore $K=\omega_2/\omega_1 = 1$ for us. Also note that, since
reference \cite{EEHG2} deals with the steady state case, the phases
of the two laser pulses are constant. This explains why only a
relative phase $\phi$ was introduced in \cite{EEHG2}. At variance,
in this paper we treat the time-dependent case, where the two laser
phases can exhibit different time variations. As a result, here we
include the phases $\phi_1$ and $\phi_2$ of both lasers.}. Using Eq.
(\ref{f0new})-(\ref{M2}) one can obtain an explicit expression for
the phase space distribution after the second stage, $f_2(\zeta'',
p'')$, which will not be reported here.
%
In order to analyze the harmonic composition of the current density
we first need to project the phase space distribution onto the real
space-time coordinates by performing an integration along $p''$.
This leads to the density distribution function $\rho$, that can be
Fourier-analyzed further to give

\begin{eqnarray}
\bar{\rho}(\Omega)= \int_{-\infty}^{\infty} dp'' d\zeta'' \exp[-i
\Omega \zeta''] f_2(\zeta'',p'')~, \cr && \label{ffft}
\end{eqnarray}
where $\Omega = \omega/\omega_l$ is the conjugate variable of
$\zeta$, whose meaning is that of normalized frequency.

The integrals in Eq. (\ref{ffft}) cannot be easily performed,
directly. As customary, one can transform the final variables
$(\zeta'', p'')$ back to the initial variables $(\zeta,p)$ and
perform the required integrations with respect to the old variables.
This allows to use the fact that $f_2(\zeta'',p'') = f_0(\zeta,p)$.
Since $d\zeta'' dp'' = d\zeta d p$ one obtains

\begin{eqnarray}
\bar{\rho}(\Omega) && = \int_{-\infty}^{\infty} dp d\zeta \exp[-i
\Omega \zeta''(\zeta,p)] f_0(\zeta,p) \cr && = \frac{N_0}{2\pi
\sigma_\zeta} \int_{-\infty}^{\infty} dp d\zeta \exp[-i \Omega
\zeta''(\zeta,p)] \exp\left[-\frac{p^2}{2}-\frac{\zeta^2}{2
\sigma_\zeta^2}\right]~, \cr && \label{ffft2}
\end{eqnarray}
where $\zeta''(\zeta,p)$ can be obtained from Eq. (\ref{M1}) and Eq.
(\ref{M2}), and reads

\begin{eqnarray}
\zeta''(\zeta,p) =&& \zeta + (B_1+B_2) p + A_1
(B_1+B_2)\sin(\zeta+\phi_1)\cr && +A_2 B_2 \sin\left(\zeta + B_1 p +
A_1 B_1 \sin(\zeta+\phi_1)+\phi_2\right) ~.\label{zeta2}
\end{eqnarray}
Substituting Eq. (\ref{zeta2}) into Eq. (\ref{ffft2}) we find,
explicitly:

\begin{eqnarray}
\bar{\rho}(\Omega) && = \frac{N_0}{2\pi \sigma_\zeta}
\int_{-\infty}^{\infty} dp d\zeta
\exp\left[-\frac{p^2}{2}-\frac{\zeta^2}{2 \sigma_\zeta^2}\right]~,
\cr && \times \exp\Bigg\{-i \Omega \Bigg[\zeta + (B_1+B_2) p + A_1
(B_1+B_2)\sin(\zeta+\phi_1)\cr && +A_2 B_2 \sin\left(\zeta + B_1 p +
A_1 B_1 \sin(\zeta+\phi_1)+\phi_2\right)\Bigg] \Bigg\}\label{ffft3}
\end{eqnarray}
The following step consists in expanding the exponential factors
containing trigonometric expressions according to\footnote{In the
following $k$ is just an index, without the meaning of wavenumber.}:

\begin{eqnarray}
\exp[-i \Omega A_1(B_1+B_2)\sin(\zeta+\phi_1)] =
\sum_{k=-\infty}^{\infty} \exp[i k (\zeta+\phi_1)] J_k\left[-\Omega
A_1(B_1+B_2)\right] \cr && \label{expan1}
\end{eqnarray}
and

\begin{eqnarray}
&&\exp\left[-i \Omega A_2 B_2 \sin\left(\zeta + B_1 p + A_1 B_1
\sin(\zeta+\phi_1)+\phi_2\right)\right] = \cr &&
\sum_{m=-\infty}^{\infty} \exp\left[i m \left(\zeta + B_1 p + A_1
B_1 \sin(\zeta+\phi_1)+\phi_2\right)\right] J_m\left[-\Omega A_2
B_2\right] ~, \label{expan2}
\end{eqnarray}
where one can still expand

\begin{eqnarray}
\exp[i m A_1 B_1 \sin(\zeta+\phi_1)] = \sum_{l=-\infty}^{\infty}
\exp[i l (\zeta+\phi_1)] J_l\left[ m A_1 B_1\right] ~.\label{expan3}
\end{eqnarray}
Assuming, for the moment, a dependence of the laser phases $\phi_1$
and $\phi_2$ on $\zeta$, and collecting terms that have a dependence
on $\zeta$ we can define

\begin{eqnarray}
\bar{f}_{\zeta}(k+l+m-\Omega)=\int_{-\infty}^{\infty} d\zeta
f_{0\zeta}(\zeta) \exp[i (k+l+m-\Omega)\zeta] \exp[i (k+l) \phi_1 +
i m \phi_2]\cr && \label{fbarz}
\end{eqnarray}
and obtain from Eq. (\ref{ffft3}):

\begin{eqnarray}
\bar{\rho}(\Omega) && = \frac{1}{\sqrt{2\pi}} \sum_{m,k,l}
\bar{f}_{\zeta}(k+l+m-\Omega) J_k\left[-\Omega A_1(B_1+B_2)\right]
J_m\left[-\Omega A_2 B_2\right] J_l\left[ m A_1 B_1\right]\cr &&
\times \int_{-\infty}^{\infty} dp \exp\left[-\frac{p^2}{2}\right]
\exp[-i \Omega (B_1+B_2)p+im B_1p]~. \label{ffft4}
\end{eqnarray}
The integration over $p$ can be carried out using

\begin{eqnarray}
\frac{1}{N_0} \int_{-\infty}^{\infty} dp \exp[-i \Omega p (B_1+B_2)
+ i m p B_1] f_{0p}(p) = \exp[(\Omega  (B_1+B_2)-m  B_1)^2/2] \cr
\label{uno}
\end{eqnarray}
which yields

\begin{eqnarray}
\bar{\rho}(\Omega) && =  \sum_{m,k,l} \bar{f}_{\zeta}(k+l+m-\Omega)
J_k\left[-\Omega A_1(B_1+B_2)\right] J_m\left[-\Omega A_2 B_2\right]
J_l\left[ m A_1 B_1\right]\cr && \times \exp[(\Omega  (B_1+B_2)-m
B_1)^2/2] ~.  \label{ffft5}
\end{eqnarray}
Setting $n=k+l$ and using

\begin{eqnarray}
J_{k+l}(\alpha+\beta) = \sum_{l=-\infty}^{\infty}
J_l(\beta)J_k(\alpha) ~,\label{due}
\end{eqnarray}
Eq. (\ref{ffft5}) can be re-written as

\begin{eqnarray}
\bar{\rho}(\Omega) && =  \sum_{m,n} \bar{f}_{\zeta}(n+m-\Omega)
J_n\left[-\Omega A_1(B_1+B_2)+m A_1 B_1\right] J_m\left[-\Omega A_2
B_2\right] \cr && \times \exp[(\Omega  (B_1+B_2)-m  B_1)^2/2] ~.
\label{ffft6}
\end{eqnarray}
We now apply the adiabatic approximation imposing that the width of
the peaks in $\bar{f}_{\zeta}$ is much narrower than the harmonic
separation $\omega_l$ between peaks. Analysis of Eq. (\ref{ffft6})
and Eq. (\ref{fbarz}) shows that due to the adiabatic approximation,
the contribution to $\bar{f}(\Omega)$ for a given value of $m+n$, is
peaked around $\Omega \simeq m+n$. This means that the terms in the
sum over $m$ in Eq. (\ref{ffft6}) can be analyzed separately for a
fixed value of $m+n$, and one obtains

\begin{eqnarray}
\bar{\rho}(\Omega,m+n) && =  \sum_{n} \bar{f}_{\zeta}(n+m-\Omega)
J_n\left[-\Omega A_1(B_1+B_2)+m A_1 B_1\right] J_m\left[-\Omega A_2
B_2\right] \cr && \times \exp[(\Omega  (B_1+B_2)-m  B_1)^2/2] ~.
\label{ffft6sep}
\end{eqnarray}
It should be remarked that due to the adiabatic approximation, and
to non-resonant behavior of Bessel functions, in Eq. (\ref{ffft6})
we can replace $\Omega$ with $m+n$ under the Bessel functions. In
this way, $\bar{f}_\zeta$ can be interpreted as the Fourier
transform of the electron bunch density. The physical meaning of all
this, is that $\bar{f}_{\zeta}$ is peaked at frequencies $\Omega$
near to multiples $n+m$ of the laser frequency. In \cite{EEHG2} it
is reported that, in order to maximize the modulus of the bunching
factor one should impose $n=\pm 1$. This can be seen directly by
inspecting the right hand side of Eq. (\ref{uno}). In fact, for
values of $\Omega$ near to $n+m$, the argument in the exponential
function can be written as $p^2 ( B_1 n +  B_2 (n+m))^2/2$. When
$n=-1$ and $m$ is positive and large for example, one sees that that
$B_1 n$ is large and negative, while $B_2 m$ is large and positive.
Therefore, $m$ can be chosen such that $ - B_1 + B_2 (m-1) \simeq
0$. This is guarantees remarkable up-frequency conversion
efficiency, almost independently on the energy spread and
constitutes one of the great advantages of the EEHG scheme.  We will
restrict our investigation to the case $n=-1$ and $m>0$, thus
obtaining

\begin{eqnarray}
\bar{\rho}(\Omega,m-1) && = \bar{f}_{\zeta}(m-1-\Omega)
J_{-1}\left[-\Omega A_1(B_1+B_2)+m A_1 B_1\right]\cr && \times
J_m\left[-\Omega A_2 B_2\right] \exp[(\Omega (B_1+B_2)-m B_1)^2/2]
~.  \label{ffft6x}
\end{eqnarray}
Note that if the laser phases would not depend on $\zeta$, which is
not true in general, one could separately calculate

\begin{eqnarray}
&&\int_{-\infty}^{\infty} d\zeta f_{0\zeta}(\zeta) \exp[i
(m-1-\Omega)\zeta] =\cr &&
\frac{1}{\sqrt{2\pi}\sigma_\zeta}\int_{-\infty}^{\infty} d\zeta
\exp\left[-\frac{\zeta^2}{2 \sigma_\zeta^2}\right]\exp[i
(m-1-\Omega)\zeta] =
\exp\left[-\frac{\sigma_\zeta^2}{2}(m-1-\Omega)^2 \right]~.\cr &&
\label{depzeta2}
\end{eqnarray}
In this case, the adiabatic approximation can be simply enforced
imposing that $\sigma_\zeta \gg 1$. Finally, it should be noted that
the initial electron density distribution and laser phases $\phi_1$
and $\phi_2$ are not only functions of $\zeta$, but also of the
transverse position $\vec{r}$. It should be understood that the
transverse direction can be factorized, which is a simplifying but
not principal assumption, and that therefore, all the expressions
above are considered valid at any fixed transverse position.

To conclude, let us consider our initial question,  whether the
general principle of the frequency multiplier chains is valid or not
for EEHG. The answer is affirmative, and can be seen by inspecting
Eq. (\ref{ffft6x}) and Eq. (\ref{fbarz}). In the case when
$\phi_1=\phi_2$ such principle can be applied strictly. In case
$\phi_1$ and $\phi_2$ differ, but are still of the same order of
magnitude, we can conclude that, since $n=-1$ and $m$ is large, only
$\phi_2$ is important and the principle is applicable with accuracy
roughly $1/N$.

\section{\label{sec:tre} Temporal quality of the seed laser beam}

Nowadays, high peak power laser systems are capable of producing
very high intensities, thus fulfilling the requirements for many
high field applications including EEHG FELs. In particular,
femtosecond laser systems have become the primary method to deal
with these applications. The reasons for this are the availability
of broadband, efficient lasing media such as titanium-doped sapphire
(Ti:Sa), and of techniques like Kerr-lens mode locking and chirp
pulse amplification (CPA). In CPA systems, light passes through a
number of optical components. Moreover, non-linear effects take
place in the amplifying medium. This can degrade the temporal
quality of the output pulse, which can be appropriately modeled in a
slowly-varying real field envelope and time-dependent carrier
frequency approximation. The time-bandwidth product constitutes a
proper measure of the departure from the ideal case, in which there
are no temporal variations of the carrier frequency. In this Section
we quantitatively describe the relation between carrier frequency
chirp and corresponding broadening of the spectrum. This leads to a
time-bandwidth product exceeding the Fourier limit.

\subsection{Pulse duration and spectral width}

For our purposes, it is convenient to consider a Gaussian pulse with
a linear frequency chirp. This choice is one of analytical
convenience only, and may be generalized. The slowly complex field
envelope is given by

\begin{eqnarray}
E(t) = A \exp\left[-\frac{t^2}{2 \tau^2}\right]\exp\left[i
\frac{\alpha t^2}{2\tau^2}\right] \label{pulsec}
\end{eqnarray}
where $\alpha$ is the chirp parameter, and the FWHM pulse duration
is related to the rms duration $\tau$ by $\Delta \tau = \sqrt{4 \ln
2} \cdot \tau$.

By Fourier transforming Eq. (\ref{pulsec}), it can be demonstrated
(see e.g. \cite{MILO}) that the spectral intensity is a Gaussian
with a FWHM given by $\Delta \omega = (\sqrt{4\ln 2}/\tau) \sqrt{1 +
\alpha^2}$. The time-bandwidth product of the pulse is therefore

\begin{eqnarray}
\Delta \omega\cdot \Delta \tau = 4 \ln 2 \cdot \sqrt{1 + \alpha^2}
\label{TBp}
\end{eqnarray}
This is larger than the time-bandwidth product of an unchirped
Gaussian pulse, which is just $4 \ln 2$ . In other words, chirping
increases the time-bandwidth product by broadening the pulse
spectrum while preserving the pulse width. Note that $\Delta \omega
\cdot \Delta \tau = 4 \ln 2$ is the smallest time-bandwidth product
for a Gaussian pulse corresponding to the transform-limit (or
bandwidth limit, or Fourier limit).

The temporal quality of the pulse can be defined by a quality factor
$M_t^2$, defined as the ratio between the time-bandwidth product for
real and transform-limited pulse. Hence, one can characterize pulse
by specifying its quality through the $M_t^2$ factor and by giving
the pulse shape. In our case of interest,  $M_t^2 = \sqrt{1 +
\alpha^2}> 1$ for Gaussian pulses with linear frequency chirp.

Finally, it should be noted that considerations analogous to those
just discussed above, can be proposed for the electron beam
microbunching. For example the current envelope of a Gaussian,
chirped electron beam can be described similarly as in Eq.
(\ref{pulsec}), with a chirp parameter $\alpha_m$. A time-bandwidth
product can be defined, and a quality factor $M_{t,m}$ can be
defined as well.

\subsection{\label{sub:con} Constraint on temporal phase variation for the output Ti:Sa
laser pulse}

There are several simplifying assumptions that will be used in our
analysis. As has been the case for the analysis presented in the
previous paragraph, we restrict our attention to a microbunched
electron beam with Gaussian shape. This is not a significant
restriction, and extensions are not difficult to consider.

We introduce the following criterion: we consider the electron beam
microbunching nearly transform-limited when the performance ratio
$M_{t,m}^{-2}$ is down not more than $1/\sqrt{2}$. For a
microbunching with Gaussian shape and linear frequency chirp, this
criterion will be satisfied under the restriction that the
microbunch chirp parameter $\alpha_m < 1$.

A specific example of a microbunched beam with Gaussian profile
could be realized in the case when EEHG scheme uses an electron
bunch with Gaussian temporal profile and a seed laser pulse with
flat-top profile in time across the duration of the electron bunch.
As demonstrated in e.g. \cite{EETOL3}, the generated bunching is not
sensitive to the peak current. Therefore,  EEHG can operate with a
nonuniform electron bunch profile. In the next paragraph we will
demonstrate that in any case, due to non-linear (self-phasing)
effects in the Ti:Sa laser system and in the post-laser optics
system, the seed laser must have flat-top profile in time with very
little temporal variation. Therefore, the model of a seed pulse with
flat-top profile and of an electron bunch with Gaussian profile is
consistent with the EEHG scheme. Now, if the phase of the seed laser
is chirped, the microbunching chirp is simply multiplied by the
frequency multiplication factor $N$. This can be seen by looking at
the harmonic contents of the current density found in Eq.
(\ref{ffft6}). That expression includes $\bar{f}_{\zeta}$, which in
the case of $\phi_1=\phi_2 = \alpha \zeta^2/(2 \sigma_\zeta^2)$ is
given by (see Eq. (\ref{fbarz})):

\begin{eqnarray}
&&\bar{f}_{\zeta}(m-1-\Omega)=\int_{-\infty}^{\infty} d\zeta
f_{0\zeta}(\zeta) \exp[i (m-1-\Omega)\zeta] \exp[-i  \phi_1 + i m
\phi_2]\cr && =
\frac{1}{\sqrt{2\pi}\sigma_\zeta}\int_{-\infty}^{\infty} d\zeta
\exp[i (m-1-\Omega)\zeta]\exp\left[-\frac{\zeta^2}{2
\sigma_\zeta^2}\right]\exp\left[i (m-1) \frac{ \alpha \zeta^2}{2
\tau_\zeta^2}\right]\cr && \label{fbarz2}
\end{eqnarray}
The last phase factor under integral shows that the laser phase is
indeed multiplied by $N=m-1$.

We will define the frequency chirp in the seed laser pulse only
across the target duration of the electron bunch, and use the same
time normalization as for the beam microbunching. The complex field
envelope of a laser pulse with stepped profile and linear frequency
chirp is given by

\begin{eqnarray}
E(\zeta) = E_0 \exp\left[i \frac{\alpha \zeta^2}{2
\sigma_\zeta^2}\right]~, \label{Et}
\end{eqnarray}
where $E_0$ is a constant. As discussed above, the frequency
multiplication yields a complex "microbunching" envelope with
carrier frequency $\omega_0 = (m-1) \omega_l$

\begin{eqnarray}
a(\zeta) = a_0 \exp\left[- \frac{\zeta^2}{2 \sigma_\zeta^2}\right]
\exp\left[i \frac{\alpha_m \zeta^2}{2 \sigma_\zeta^2}\right]
\label{Et}
\end{eqnarray}
where $\rho_0$ is a constant, and $\alpha_m = N\alpha$ is the
microbunching chirp parameter. Note that what we loosely defined as
"microbunching" is, more formally, the slowly-varying amplitude of
the electron density modulation with carrier frequency $\Omega=N$.
It follows from the previous analysis that the EEHG scheme can
produce nearly transform-limited microbunching only under the
restriction $\alpha_m \lesssim 1$, meaning that the laser chirp
parameter must obey $\alpha \lesssim 1/N$. The EEHG seed laser is
assumed to be a Ti:Sa laser. The actual seed laser beam consists in
the third or in the fourth harmonic of the Ti:Sa laser beam.
Usually, laser frequency multipliers are based on the use of Beta
Barium Borate (BBO) crystals. The effect of frequency multiplication
on phase variation amounts again to multiplication of the phase
variations. Therefore we may say that when we study constraints on
the performance of Ti:Sa seed laser for EEHG schemes, the total
frequency multiplication chain consists of two stages. The first
stage is the BBO crystals with a frequency multiplication factor
$N_1 =3$ (or $N_1 =4$). The second stage is the EEHG setup itself,
with frequency multiplication factor up to $N_2 \sim 270$ (or $N_2
\sim 200$).  If the final required output radiation is around
wavelengths of $1$ nm, the total frequency multiplication factor $N
= N_1 N_2$ is about $N \sim 800$.

From the previously discussed condition $\alpha \lesssim 1/N$ it can
be seen that the Ti:Sa laser produces nearly transform-limited
microbunching at wavelengths around $1$ nm only when the laser chirp
parameter $\alpha \lesssim 10^{-3}$.  Thus, for most purposes, if
the total multiplication factor is around $800$ or exceeds it, we
may formulate the constraint on the Ti:Sa laser quality by requiring
a quality factor $M_t^2$ departing from unity of no more than about
$10^{-6}$.

One can think that the above-discussed constraints on seed laser may
be true only for the particular case of EEHG. However, we can show
that these constraints are actually of more general validity.  For
example, HGHG schemes can produce nearly transform-limited radiation
spanning down to wavelengths of $1$ nm only under the same
restrictions on temporal quality of the seed Ti:Sa laser. The key
advantage of the EEHG scheme is that the amplitude of the  achieved
microbunching factor slowly decays with increasing harmonic number
and that, consequently, generation of coherent soft X-ray emission
within a single upshift stage becomes possible \cite{EEHG1,EEHG2}.
However, considering constraints on the seed laser $M_t^2$ factor,
all harmonic generation schemes are similar, and must obey the
universal result

\begin{eqnarray}
M_t^2-1 \lesssim \frac{1}{N^2}~. \label{unires}
\end{eqnarray}
The requirement in the inequality (\ref{unires}) can be somehow
relaxed if the requirement of near-Fourier limit is relaxed as well.
For example, the operation of a EEHG FEL is characterized by two
microbunch bandwidth scales of interest.   One is associated with
inverse electron bunch duration $\Delta \omega_b = 1/\tau_b$,
$\tau_b$ being the electron bunch duration. The other is the FEL
amplification bandwidth $\Delta \omega_a$. One can relax the
requirement of near-Fourier limit substituting it by the requirement
to achieve an output radiation bandwidth narrower than the SASE
bandwidth $\Delta \omega_a$. On the one hand, the product of bunch
duration by amplification bandwidth can be estimated in the order of
$\tau_b \Delta \omega_a \sim 10^2$ in the soft X-ray wavelength
range. On the other hand, the FEL radiation bandwidth broadening due
to the effect of linear frequency chirp is about $\Delta \omega \sim
|\alpha_m|/\tau_b$. Therefore, in the case when

\begin{eqnarray}
|\alpha_m| >   (\tau_b \Delta \omega_a) \sim 10^2 ~,\label{alpham}
\label{alpham}
\end{eqnarray}
the output signal has a bandwidth larger than the SASE bandwidth,
and harmonic generation techniques have no practical applications.
However, if, for example, we have a microbunching chirp parameter
$|\alpha_m| \sim 10$, the effective radiation bandwidth becomes ten
times narrower than the SASE bandwidth, although is ten times wider
compared to the ideal transform-limited bandwidth. Following this
discussion, a weaker constraint on the temporal quality factor of
seed laser is $M_t^2 - 1 < 10^2/N^2$. For a Ti:Sa laser seed and a
radiation wavelength of $1$ nm it is possible to discuss about
harmonic generation techniques applications only when $M_t^2 - 1 <
10^{-4}$.

To complete the picture, we should note that an alternative method
to harmonic generation setups, called self-seeding
\cite{SELF,TREU,MARI}, is available, and allows for the generation
of temporally coherent radiation in XFELs. A self-seeded soft X-ray
FEL consists of two undulators separated by a monochromator
installed within a magnetic chicane. The remarkable temporal quality
of the output radiation and the wavelength tunability of
self-seeding schemes has stimulated interest in using this technique
to generate nearly transform-limited soft- X-ray pulses. A project
of self-seeding schemes with grating monochromator is now under
development at LCLS II  \cite{LCLS2}-\cite{FENG}. EEHG output will
compete with self-seeding output only when the temporal quality of
the seed laser beam obeys the mores stringent requirement
(\ref{unires}).

\subsection{\label{sub:self} Self-phasing and constraints on field amplitude variation}

The seeding pulse from the Ti:Sa laser must necessarily propagate
through vacuum window and BBO crystals without experiencing temporal
phase distortions. Above a power density of $1 \mathrm{GW/cm^2}$,
the refractive index $n$ becomes intensity-dependent according to
the well-known expression

\begin{eqnarray}
n = n_0 + n_2 I ~, \label{refrin}
\end{eqnarray}
where $n_0$ is the index of refraction at low intensity and $I$ is
the laser intensity. Due to temporal variations of the laser pulse
intensity, the pulse phase will then be distorted according to
\cite{MILO}

\begin{eqnarray}
B  = \frac{2\pi}{\lambda} \int_0^{L} dz n_2 I ~.\label{B}
\end{eqnarray}
Here $\lambda$ is the laser wavelength, and $B$ represents the
amount of phase distortions accumulated by the pulse over a length
$L$. The dimensionless $B$ parameter, also known as $B$ integral, is
often used as a measure of the strength of nonlinear effects due to
the non-linear refractive index $n_2 I$. Field intensities,
propagation distances, and values of $n_2 I$ such that $B> 1$
generally yield significant nonlinear effects, including self-phase
modulation. Usually, in laser optics, when $B < 0.5$ pulse
distortions should not be a problem.

Let us consider an optical setup behind the Ti:Sa laser with $B \sim
0.5$. In order to have minimal FEL output spectral broadening, the
seed laser must have flat-top profile in time with very little
temporal variation. The intensity variation must satisfy

\begin{eqnarray}
\frac{\Delta I}{I} <  \frac{2}{N} \label{DII}
\end{eqnarray}
For $1$ nm wavelength mode of operation  $N \sim 800$, and in the
case of near transform-limited FEL output pulse, the  intensity  of
Ti:Sa laser pulse must be controlled to about $0.3 \%$ across the
target duration of the electron bunch.

\section{\label{sec:quattro} Spatial quality of seed laser beam}

In the last section we considered part of the constraints on the
performance required for EEHG seed lasers. In particular, our
discussion has been restricted to the temporal quality of laser
beams. The former restriction allows one to obtain results which
depend on the frequency multiplication factor only, so that the
treatment discussed above applies not only to EEHG schemes, but to
more general cases as well. In this section we discuss, instead, the
influence of errors on the wavefront of the seed laser beam. A
general principle discussed before states that the effect of
frequency multiplication by a factor $N$ is to multiply the phase
variation in time by $N$. The same principle holds when dealing with
phase variations in space. If the wavefront of the UV seed laser
exhibits errors, the errors of the microbunching wavefront are
multiplied by the frequency multiplication factor. This can be seen
with an analysis similar to that in paragraph \ref{sub:con}, based
on the results in Section \ref{sec:due}, which led to Eq.
(\ref{fbarz2}). However, now, the phase variations are to be
considered as a function of spatial coordinates. In the case of
variation in time, the temporal quality of the output FEL radiation
is a replica of the temporal quality of the microbunching input. It
seems natural to use the same principle for characterizing the
spatial quality of the output FEL radiation. However, this cannot be
done. The reason is that the transverse coherence of FEL radiation
is settled without laser seeding.  This is due to the transverse
eigenmode selection mechanism: only the ground eigenmode survives at
the end of the amplification process. It follows that the
microbunching wavefront errors do not affect the spatial quality of
the output radiation. They only affect the input signal value. The
description of the influence of phase errors depends in detail on
the harmonic generation process. For example, in the case of HGHG,
the seed laser directly produces microbunching in the first cascade
only, which is characterized by a relatively small frequency
multiplication factor $N < 5$.  In EEHG schemes instead, the
generation of coherent radiation in the soft X-ray wavelength range
should be achieved with a single upshift stage using a UV ( $200$ nm
or $270$ nm) laser beam. In this case the frequency multiplication
factor amounts to about  $N \sim 200$. Consequently, the EEHG
technique is much more sensitive to laser wavefront errors. This
disadvantage is actually related to the key EEHG advantage, that is
to allow for high frequency multiplication numbers within a single,
compact scheme.

To understand the effects of wavefront errors we shall use an
analogy between time and space. This analogy suggests the
possibility of simply translating the effects related to phase
perturbation in time into effects related to wavefront perturbations
as shown in Table \ref{tt1}.

\begin{table}
\caption{Analogy between temporal and spatial characteristics}

\begin{small}\begin{tabular}{ p{0.5\textwidth} p{0.5\textwidth} }
\hline Temporal (pulse)& ~ Spatial (beam) \\ \hline
transform-limited pulse      & diffraction-limited beam     \\
temporal frequency           &  spatial frequency  \\
bandwidth of amplification   &  bandwidth of amplification \\
temporal frequency shift (temporal linear phase chirp)    &
wavefront tilt
(spatial linear phase chirp)     \\
linear temporal frequency chirp (temporal quadratic phase chirp) &
defocusing aberration
(spatial quadratic phase chirp) \\
nonlinear temporal frequency chirps  &  high order wavefront
aberrations \\
phase fluctuations in time  & chaotic phase variation across the
beam\\
\hline
\end{tabular}\end{small}
\label{tt1}
\end{table}
We defined the ideal seed pulse as a transform-limited pulse i.e. a
pulse without phase variations in time. The space-domain analog of a
transform-limited pulse is a diffraction-limited beam, i.e. a beam
without phase variations in space. From this definition follows that
a beam can be diffraction-limited only at its waist, where it takes
on the minimum possible product between size and spatial frequency
bandwidth. In fact, beam propagation leads to a beam broadening and
to a spatial quadratic phase chirp. Since the ideal seed laser beam
is characterized by microbunching wavefront without phase variation
across the electron beam, it follows that the seed laser beam must
be diffraction-limited at its waist, which must be placed in the
middle of the modulator undulator.

Simple physical considerations directly lead to a crude
approximation for the amplification bandwidth. As already discussed
in paragraph \ref{sub:con}, in the time domain the amplification
bandwidth is about two order of magnitudes larger than
transform-limited bandwidth:

\begin{eqnarray}
\tau_b \Delta \omega_a \sim 10^2 ~. \label{taub}
\end{eqnarray}
This fact has some interesting consequences. Suppose that we
consider microbunching with linear phase chirp in time, which is
actually equivalent to a shift of the signal frequency. In the case
when the shift is smaller than the amplification bandwidth, the
temporal quality and the output power of the radiation pulse are not
changed. At variance, microbunching with nonlinear phase chirp leads
to a spectral broadening of the output radiation and, consequently,
to degradation of the temporal quality. However, in the case when
the broadening is smaller than the amplification bandwidth, the
output power is not suppressed. The situation is quite different
when considering the spatial domain. In fact, qualitatively, the
spatial frequency amplification bandwidth and the
diffraction-limited bandwidth are the same, so that any  shift or
broadening of the spatial frequency spectrum immediately leads to
input signal suppression.

Let us study the discrepancy between the direction of the electron
motion and the normal to the microbunching wavefront. In the case
when the discrepancy between these two directions is larger than the
FEL angular amplification bandwidth the input signal is
exponentially suppressed. Let us assume that the spatial profile of
the microbunching is close to that of the electron beam, and is
characterized by a Gaussian shape with standard deviation
$\sigma_b$. The FEL angular amplification bandwidth can then be
estimated as $\Delta \theta_a \sim (k \sigma_b)^{-1}$, where $k$ is
wavenumber at the target harmonic.

One can then estimate the angular spectrum of e.g. the LCLS output
for the wavelength of $1.5$ nm.  The transverse distribution of the
electron beam is described by $\sigma_b \sim 30 \mu$m, and our
estimations give $\Delta \theta_a \sim  8 \mu$rad. The angular
amplification bandwidth corresponds to the HWHM of the FEL output
angular distribution. Results of numerical simulations, confirmed by
experimental results, give an angular distribution of the radiation
intensity with HWHM $\sim 10 \mu$rad. From these numbers one can see
that the above approach  provides an adequate description, at least
in the wavelength range around $1$ nm. The value $\Delta \theta_a$
can subsequently be used to estimate the maximum angular error
allowed between the normal to the laser beam wavefront (at its
waist) and the direction of the electron beam motion in the
modulator undulator. It follows from the previous reasoning that in
the case of radiation wavelength around $1$ nm we find an alignment
tolerance of about $10 \mu$rad.

The wavefront tilting is a relatively simple (first order)
geometrical distortion and its measure is simply an angle, which is
the same for the microbunching wavefront and for the laser beam
wavefront. The width of the seed laser beam at its waist can be much
larger than the width of the electron beam, but the tilt is
completely characterized by such angle only. There are several
criteria to analyze the performance of laser system to higher order
aberrations.  To characterize the spatial quality of the laser beam,
we will use the Strehl ratio $S$, usually defined\footnote{With this
definition, the Strehl ratio is related to the transverse $M^2$
parameter by $S=1/M^2$} as:

\begin{eqnarray}
S=\frac{\max[|FT\{E(x,y)\exp[i\phi(x,y)]\}|^2]}{\max[|FT[E(x,y)]|^2]}~,
\label{Sratio} \end{eqnarray}
where "FT" indicates the 2D spatial Fourier transform operation,
$E(x,y)$ is the ideal wave amplitude, and $\phi(x,y)$ is the phase
aberration. The Strehl ratio $S$ becomes an important figure of
merit from the viewpoint of seeding evaluation.

Let us consider the practical situation in which both laser and
electron beams are characterized by a Gaussian shape, and in which
the width of the laser beam at its waist is much larger than the
width of the electron beam. With this assumption, within the
electron beam, at the laser beam waist in the plane $z = 0$  we have
asymptotically $E(x,y,0) = \mathrm{const}\cdot \exp[i\phi(x,y)]$ ,
where $E(x,y,0)$ is the wave amplitude, and $\phi(x,y)$ describes
phase aberrations. For our purposes it is interesting to consider
the Gaussian-weighted Strehl ratio $S$

\begin{eqnarray}
S =  \left|\left<\exp[i \phi(x,y)]\right>\right|^2 ~,
\label{Sweight}
\end{eqnarray}
where

\begin{eqnarray}
\left<\exp[i\phi(x,y)]\right> =  (2\pi \sigma_p^2)^{-1} \int dx dy
\exp\left[-\frac{x^2+y^2}{2 \sigma_p^2}\right]\exp[i\phi(x,y)]~.
\label{aveX}
\end{eqnarray}
Here $\sigma_p$ is a Gaussian parameter of the same order of
magnitude of the rms width of the electron beam, $\sigma_b$. If the
phase is sufficiently small to accurately replace $\exp[i\phi]$ with
$1+ i\phi -\phi^2/2$, one obtains

\begin{eqnarray}
S = 1 - \sigma_\phi^2 ~, \label{S2}
\end{eqnarray}
where

\begin{eqnarray}
\sigma_\phi^2 = <\phi^2> -  <\phi>^2 \label{sigphi}
\end{eqnarray}
is the variance of the phase aberration weighted across a
Gaussian-amplitude pupil.  To be more specific, we define the
average of $\phi(x,y)$ across the pupil as

\begin{eqnarray}
<\phi> = (2 \pi \sigma_p^2)^{-1} \int dx dy \exp\left[- \frac{x^2 +
y^2}{2 \sigma_p^2}\right] \phi(x,y) ~,\label{avephi2}
\end{eqnarray}
and, likewise, the average of the square of $\phi(x,y)$  as

\begin{eqnarray}
<\phi^2> =  (2\pi\sigma_p^2)^{-1} \int dx dy \exp\left[-\frac{x^2 +
y^2)}{2\sigma_p^2}\right] \phi^2 ~. \label{phi2ave2}
\end{eqnarray}
It follows that if the root-mean-square variations of the wavefront
are of the order of a tenth of the wavelength only, we obtain a
Strehl ratio of $0.6$.

Let us now discuss the spatial quality of the microbunching
wavefront. The interesting value to know for EEHG operation is the
input coupling factor between the microbunching and the ground
eigenmode of the FEL amplifier.  Let us consider the amplitude of
the electron density modulation  at the carrier frequency $\omega_0
= (m-1) \omega_l$ :

\begin{eqnarray}
{\rho}(x,y,t) =  a(x,y,t)\exp[i (m-1) \omega_l t]~. \label{tilderho}
\end{eqnarray}
In ideal case, the electron density modulation exhibits a plane
wavefront and a Gaussian shape across the electron beam:

\begin{eqnarray}
a(x,y,t) = a_0(t)\exp\left[-\frac{x^2+y^2}{2\sigma_b^2}\right]~.
\label{bxyt} \end{eqnarray}
In such ideal case, the input coupling factor is therefore

\begin{eqnarray}
C = \int dx dy \exp
\left[-\frac{x^2+y^2}{2\sigma_b^2}\right]\Psi(x,y)~, \label{CCCC}
\end{eqnarray}
where $\Psi(x,y)$ is the field distribution of the ground eigenmode.
In the high gain linear regime, the FEL output radiation power
scales as

\begin{eqnarray}
W_\mathrm{output} \sim |C|^2 ~. \label{Wout}
\end{eqnarray}
In the case of a non-ideal microbunching wavefront, expressions for
$a(x,y,t)$ and for the input coupling factor respectively transform
to:

\begin{eqnarray}
a(x,y,t) = a_0(t)\exp[i\phi_m(x,y)]
\exp\left[-\frac{x^2+y^2}{2\sigma_b^2}\right]~ , \label{bnonid}
\end{eqnarray}
and

\begin{eqnarray}
C = \int dx dy~ a(x,y,t) \Psi(x,y)  ~, \label{Cnonide}
\end{eqnarray}

where $\phi_m(x,y)$ is the microbunching phase aberration. The ratio
of the output power for the case including microbunching wavefront
errors to the output power for the case of a plane microbunching
wavefront is a simple and convenient measure of the departure from
the ideal situation. In our case this ratio is simply

\begin{eqnarray}
\frac{W_{\mathrm{nonideal}}}{W_{\mathrm{ideal}}}
=\frac{|C_\mathrm{nonideal}|^2}{|C_\mathrm{ideal}|^2} ~.
\label{ratiomer}
\end{eqnarray}
Various  approximations can be invoked. One of the simplest is to
use the following expression for the ground FEL eigenfunction

\begin{eqnarray}
\Psi(x,y) \sim  \exp\left[-\frac{x^2+y^2}{2\sigma_b^2}\right] ~.
\label{eigenmode}
\end{eqnarray}
With this approximation it can be shown that

\begin{eqnarray}
\frac{|C_\mathrm{nonideal}|^2}{|C_\mathrm{ideal}|^2} = 1-
\sigma_\phi^2 ~, \label{Cration}
\end{eqnarray}
where $\sigma_\phi^2$ is the variance of the microbunching phase
aberration across the Gaussian-weighted pupil with

\begin{eqnarray}
\sigma_p  = \frac{\sigma_b}{\sqrt{2}}~ . \label{sigp2}
\end{eqnarray}
If we now look at the ratio of the power values at the FEL exit with
microbunch wavefront distortions and without distortions, we see
that such ratio corresponds to the already introduced laser Strehl
ratio, Eq. \ref{S2}. More in general, we have the same definition
given in Eq. (\ref{Sweight}), where the phase $\phi$ under the
integral is now defined as the phase on the microbunching wavefront
$\phi_m$.

Finally, we calculate the relation between the phase distortions of
the laser beam and the phase distortions of the microbunching. We
have concluded from our theoretical analysis in Section
\ref{sec:due}, that if the wavefront of the seed laser beam in the
waist plane exhibits errors, the errors of the microbunching
wavefront are multiplied by the frequency multiplication factor $N$.
Therefore we have

\begin{eqnarray}
(\sigma_\phi)_\mathrm{laser} =
\frac{1}{N}(\sigma_\phi)_\mathrm{microbunch}~, \label{sigphimicro}
\end{eqnarray}
which yields

\begin{eqnarray}
1-S_\mathrm{laser} = \frac{1}{N^2} [1-S_\mathrm{microbunch}] ~.
\label{1ms}
\end{eqnarray}
For EEHG schemes, $[1-S_\mathrm{microbunch}]$ must be kept below
$0.4$, corresponding to microbunching wavefront distortions of
$\lambda/10$. This corresponds to a UV laser Strehl ratio
$S_\mathrm{laser} > 0.99999$ at the target wavelength of $1$ nm.

In order to experimentally investigate the effects of laser
wavefront errors on the FEL amplification process, one should
perform  direct measurements of the laser beam wavefront using, for
example, a Hartmann sensor. Usually, measurements of the spatial
quality of the output laser beam with a Hartmann sensor give the
near-field wavefront characteristics. The knowledge of the spatial
phase and amplitude in a particular plane opens the possibility of
calculating, by Fresnel propagation,  the phase and amplitude in any
other plane for a freely propagating laser beam, and in particular
allows to recover results in the middle plane of the modulator
undulator. Applying the definition of the Gaussian-weighted Strehl
ratio  in Eq. (\ref{Sweight}) with $\sigma_p =\sigma_b/\sqrt{2}$
leads to the value which needs to be compared with constraint

\begin{eqnarray}
1-  S  <  \frac{0.4}{N^2}~. \label{last}
\end{eqnarray}
The arguments discussed above seem to be strong enough to suggest
that EEHG FEL schemes for reaching frequency multiplication factor
of $N$ will not work when the difference of the above-defined laser
Strehl ratio from the unity does not satisfy the inequality in
(\ref{last}). This conclusion for the spatial domain contrasts with
that in the time domain, where the phase distortions lead to
spectral broadening but do not have an impact on the FEL output
power.

\section{\label{cinque} Conclusions}

It is very desirable to have a way to model the performance of EEHG
FEL with high frequency multiplication factor. Such modeling would
naturally start with the Ti:Sa laser system. Calculations would
involve the knowledge of the temporal and spatial properties of the
Ti:Sa laser source itself together with laser field propagation
through the optical components used in the EEHG beamline. Most of
our calculations are, in principle, straightforward applications of
conventional laser optics and general theory of frequency multiplier
chains. Our paper provides physical understanding of the  laser
seeding setup and we expect it to be useful for practical
estimations, especially at the design stage of the experiment.
Detailed EEHG mechanism is so complicated that we cannot accurately
determine the EEHG output by analytical methods. However, a definite
relation between quality of the input signal and EEHG FEL output can
be worked out without any knowledge about the EEHG internal
machinery.

\section{Acknowledgements}

We are grateful to Massimo Altarelli, Reinhard Brinkmann, Serguei
Molodtsov and Edgar Weckert for their support and their interest
during the compilation of this work.


\begin{thebibliography}{99}

\bibitem{HGHG1} L. -H. Yu. Phys. Rev. A 44, 5178 (1991).

\bibitem{HGHG2} L. -H. Yu, I. Ben-Zvi, Nucl. Instrum. Methods A
393, 96-99 (1997).

\bibitem{EEHG1} G. Stupakov, Phys. Rev. Lett. 102, 074801 (2009).

\bibitem{EEHG2} D. Xiang and G. Stupakov, Phys. Rev ST AB 12, 030702
(2009).

\bibitem{LCLS2E} D. Xiang and G. Stupakov "Echo-seeding options for LCLS-II", TUPB13, Proceedings of FEL 2010, Malmo, Sweden
(2010).

\bibitem{SWIEEHG} E. Prat and S. Reiche, "EEHG seeding design for SwissFEL", TUPA25, Proceedings of FEL 2011, Shanghai,
China (2011).

\bibitem{FLEEHG} K. E. Hacker, et al., Echo-seeding experiment at FLASH in 2012", TUPB10, Proceedings of FEL 2011, Shanghai,
China (2011).

\bibitem{EETOL1} D. Xiang and G. Stupakov "Tolerance study for the EEHG Laser", WE5RFP044, Proceedings of PAC09, Vancouver, BC,
Canada (2009).

\bibitem{EETOL2} Z. Huang, et al., "Effect of energy chirp on EEHG Lasers", MOPC45, Proceedings of FEL 2009, Liverpool, UK
(2009).

\bibitem{EETOL3} G. Penn and M. Reinsch, Journal of Modern Optics,
1-15 (2011).

\bibitem{RATN} D. Ratner,
https://sites.google.com/a/lbl.gov/
realizing-the-potential-of-seeded-fels-in-the-soft-x-ray-regime-workshop/talks,
Workshop on Realizing the Potential of Seeded FELs in the Soft X-Ray
Regime, Berkeley, CA, USA (2011) and following discussion sessions.

\bibitem{WORK1} K. Hacker, "EEHG at FLASH 2012 and beyond",
https://indico.desy.de/conferenceDisplay.py?ovw=True$\setminus
\&$confId=4736, FLASH Accelerator Workshop, Hamburg, Germany (2011)

\bibitem{WORK2} K. Hacker,
https://sites.google.com/a/lbl.gov/realizing-the-potential-of-seeded-fels
-in-the-soft-x-ray-regime-workshop/talks, Workshop on Realizing the
Potential of Seeded FELs in the Soft X-Ray Regime, Berkeley, CA, USA
(2011) and following discussion sessions.

\bibitem{ROBI} W. P. Robins, "Phase Noise in Signal Sources", IEEE
Telecommunication Series, vol 9., Peter Peregrinus Ltd. (1982).


\bibitem{MILO} P. Milonni and J. Eberly, "Laser Physics", Wiley and
Sons, (2010).


\bibitem{SELF} J. Feldhaus et al., Optics. Comm. 140, 341 (1997).

\bibitem{TREU} R. Treusch, W. Brefeld, J. Feldhaus and U. Hahn, HASILAB
Ann. report 2001 "The seeding project for the FEL in TTF phase II"

\bibitem{MARI} A. Marinelli et al., Proceedings of the FEL Conference 2008, MOPPH009 (2008).

\bibitem{LCLS2} The LCLS-II Conceptual Design Report, Stanford,
https:// slacportal.slac.stanford.edu/
sites/lcls$\_$public/lcls$\_$ii/Pages/default.aspx (2011).

\bibitem{WU} J. Wu et al., "Staged self-seeding scheme for narrow
bandwidth , ultra-short X-ray harmonic generation free electron
laser at LCLS", proceedings of 2010 FEL conference, Malmo, Sweden,
(2010).

\bibitem{FENG} Y. Feng et al., "Optics for self-seeding soft x-ray FEL
undulators", proceedings of 2010 FEL conference, Malmo, Sweden,
(2010).




%
%
%
%
%
%
%
%
%
%
%
%
%
%
%
%
%
%
%
%
%
%
%
%
%
%
%
%
%

\end{thebibliography}
\end{document}